\documentclass[prl,showpacs,superscriptaddress,twocolumn]{revtex4}

\usepackage{amsmath,bm}
\usepackage{graphicx}
\usepackage{epstopdf}

\usepackage{color}
\begin{document}

\newcommand{\vk}{{\vec k}}
\newcommand{\vK}{{\vec K}}
\newcommand{\vb}{{\vec b}}
\newcommand{{\vp}}{{\vec p}}
\newcommand{{\vq}}{{\vec q}}
\newcommand{\vQ}{{\vec Q}}
\newcommand{\vx}{{\vec x}}
\newcommand{\beq}{\begin{equation}}
\newcommand{\eeq}{\end{equation}}
\newcommand{\half}{{\textstyle \frac{1}{2}}}
\newcommand{\gton}{\stackrel{>}{\sim}}
\newcommand{\lton}{\mathrel{\lower.9ex \hbox{$\stackrel{\displaystyle<}{\sim}$}}}
\newcommand{\ee}{\end{equation}}
\newcommand{\ben}{\begin{enumerate}}
\newcommand{\een}{\end{enumerate}}
\newcommand{\bit}{\begin{itemize}}
\newcommand{\eit}{\end{itemize}}
\newcommand{\bc}{\begin{center}}
\newcommand{\ec}{\end{center}}
\newcommand{\bea}{\begin{eqnarray}}
\newcommand{\eea}{\end{eqnarray}}

\newcommand{\beqar}{\begin{eqnarray}}
\newcommand{\eeqar}[1]{\label{#1} \end{eqnarray}}
\newcommand{\pleft}{\stackrel{\leftarrow}{\partial}}
\newcommand{\pright}{\stackrel{\rightarrow}{\partial}}

\newcommand{\eq}[1]{Eq.~(\ref{#1})}
\newcommand{\fig}[1]{Fig.~\ref{#1}}
\newcommand{\eff}{ef\!f}
\newcommand{\alphas}{\alpha_s}

\renewcommand{\topfraction}{0.85}
\renewcommand{\textfraction}{0.1}
\renewcommand{\floatpagefraction}{0.75}
\renewcommand\thesection{\arabic {section}}
\setcounter{section}{1}

\title{ Probing cold nuclear matter effects with the productions
of isolated-$\gamma$ and $\gamma$+jet
in p$+$Pb collisions at $\sqrt{s_{NN}}=$ 8.16 TeV }

\date{\today  \hspace{1ex}}
\author{Guo-Yang Ma}
\affiliation{Key Laboratory of Quark $\&$ Lepton Physics (MOE) and Institute of Particle Physics,
 Central China Normal University, Wuhan 430079, China}

\author{Wei Dai\footnote{weidai@cug.edu.cn}}
 \affiliation{School of Mathematics and Physics, China University of Geosciences, Wuhan 430079, China}
%\affiliation{Key Laboratory of Quark \& Lepton Physics (MOE) and Institute of Particle Physics,
% Central China Normal University, Wuhan 430079, China}

\author{Ben-Wei Zhang\footnote{bwzhang@mail.ccnu.edu.cn}}
%\email{bwzhang@mail.ccnu.edu.cn}
\affiliation{Key Laboratory of Quark $\&$ Lepton Physics (MOE) and Institute of Particle Physics,
 Central China Normal University, Wuhan 430079, China}

\begin{abstract}

We investigate the cold nuclear matter (CNM) effects
on the productions of the isolated prompt photon and $\gamma+$jet
in proton-lead collisions at $\rm 8.16$ TeV under the next-to-leading
order (NLO) perturbative quantum chromodynamics calculations with
four parametrizations for nuclear parton distribution functions (nPDFs),
{\it i.e.} DSSZ, EPPS16, nCTEQ15, nIMParton.
Our theoretical calculations provide good descriptions of pp baseline
in the ATLAS collaboration and make predictions for future
experimental results at $\rm p$+$\rm Pb$ collisions. We calculate the dependence of
the nuclear modification factor of isolated prompt photon on transverse momentum $\rm p_T^{\gamma}$
and pseudo-rapidity $\eta^{\gamma}$ at very forward and backward rapidity regions, and demonstrate that
the forward-to-backward yield asymmetries $\rm Y_{\rm pPb}^{\rm asym}$ as a function of $\rm p_T^{\gamma}$ with different nPDFs parametrizations have diverse behaviors.
Furthermore, the nuclear modification factor of isolated-$\gamma+$jet $\rm R_{\rm pPb}^{\gamma Jet}$ as a function of $\gamma+$jet's pseudo-rapidity
$\eta_{\gamma \rm Jet}=\frac{1}{2}(\eta_{\gamma}+\eta_{\rm Jet})$ at different
average transverse momentum $\rm p_T^{\rm avg}=\frac{1}{2}(\rm p_T^{\gamma}+\rm p_T^{\rm Jet})$ has been discussed,
which can facilitate a tomography study of  CNM effects with precise locations in a rather wide kinematic region by varying the transverse momenta and rapidities of both isolated photon and jet in p+A collisions.

\end{abstract}

\pacs{12.38.Mh; 25.75.-q; 13.85.Ni}\

\maketitle

\clearpage
\section{1 Introduction}
\label{Introduction}
In high energy nuclear physics, productions of prompt photon and photon associated jet
with high transverse momentum are valuable observations
of the short-distance dynamics of quarks and gluons~\cite{Catani:2002ny}.
Since prompt photon is precisely calculable by perturbative quantum chromodynamics (pQCD)
at higher order corrections and carries no color charge like other gauge bosons,
it has been widely regarded as an optimal probe of the initial state of the collisions\cite{Janus:2018jto,Ru:2014yma,Ru:2015pfa,Ru:2016ifu,Ru:2016wfx},
as well as an excellent tag of inversive parton (jet) to quantify the mechanisms of
jet quenching in ultra-relativistic heavy ion collisions~\cite{Chatrchyan:2012vq,Wang:1996pe,Wang:1996yh,Albacete:2017qng,Dai:2013xca,Helenius:2014qla,Goharipour:2017uic,Goharipour:2018sip}.
%Meanwhile the cold nuclear matter(CNM) effects need to be considered
%in the heavy ion collisions(HIC) as an reliable benchmark to the final state medium effects.
In the past few years, ATLAS~\cite{Aaboud:2016sdm,Aaboud:2017cbm,ATLAS:2017ojy} and CMS
\cite{Sirunyan:2018gro,Sirunyan:2018qec,Sirunyan:2017qhf} collaborations have made lots of measurements on
isolated prompt photon and $\gamma$+jet productions in proton-proton,
proton-nucleus and nucleus-nucleus collisions.
The isolated photon's nuclear modification factor depending on photon transverse energy $\rm E_T^{\gamma}$
measured in most central PbPb collisions has large uncertainty, it also shows centrality dependence
at limited $\rm E_T^{\gamma}$ intervals~\cite{Chatrchyan:2012vq}.
And $\gamma$+jet events have been discussed to extend the study on the tomography of
quark-gluon plasma(QGP) created in heavy ion collisions.
For instance, the distribution of photon plus jet transverse momentum imbalance $(\rm x_{j\gamma}=p_T^{jet}/p_T^{\gamma})$ indicates
that a larger part of jet's initial energy would be damped and deposited in most central Pb$+$Pb collisions\cite{Sirunyan:2017qhf}.   In this study we may employ the productions of isolated photon and
photon+jet in p+A collisions to probe the initial-state cold nuclear matter (CNM) effects.

%At leading order, prompt photons are produced back-to-back with an associated jet
%carrying the same amount of transverse momentum in the hard scattering.
%Therefore, photon+jet production has been hailed as the "golden channel"
%to investigate the final state jet quenching effect in A+A collisions,
In elementary hadron-hadron collisions, with the perturbative QCD (pQCD), the cross section of leading particle (and jet) in general could be expressed as an convolution of
the parton distribution functions (PDFs), and the hard partonic cross section, and the fragmentation functions (FFs) if applicable.
The parton distribution function (PDF) for a parton i from the free proton $\rm (f_i^p(x,Q^2))$ is of nonperturbative property in the frame of
the QCD collinear factorization theorem~\cite{Collins:1989gx,Brock:1994er} and its evolution in the scale $\rm Q^2$
can be depicted as the DGLAP equations \cite{Dokshitzer:1977sg,Gribov:1972ri,Altarelli:1977zs}.
In p+A collisions, PDFs in nuclear environment should be modified due to different CNM effects, such as
shadowing, anti-shadowing, EMC effect and Fermi motion {\it etc.} \cite{deFlorian:2003qf}.  It is expected that the QCD factorization theorem may hold for nuclei as a good approximation, and we can replace
the PDF in a free proton  $\rm (f_i^p(x,Q^2))$ with the nuclear PDF
$\rm (f_i^A(x,Q^2))$ to effectively include different CNM effects to study hard processes in p+A collisions.

In past three decades, our understanding on the global fits of nuclear PDFs (nPDFs)
have been regularly enriched by the growing experimental results of
the fixed-target deeply inelastic scattering (DIS) and low-mass Drell-Yan (DY) dilepton measurements
\cite{Eskola:1998df,Eskola:2012rg,Paukkunen:2014nqa} and
%\cite{Eskola:1998df}\cite{Eskola:2012rg}\cite{Paukkunen:2014nqa}
the theoretical predictions from leading order (LO) up to next-to-next-to-leading order (NNLO)
\cite{Pumplin:2002vw,Gao:2013xoa,Ball:2014uwa,Harland-Lang:2014zoa}.
In the DSSZ framework~\cite{deFlorian:2011fp}, the global analysis for the nPDFs is presented as the ratio of
parton distributions in a proton of a nucleus and in the free proton,
$\rm R_i^A(x,Q^2)=f_i^A(x,Q^2)/f_i^p(x,Q^2)$ evolving at initial scale $\rm Q_0 = 1~GeV$.
The DSSZ analysis not only uses the $l^\pm/\mu-$DIS data sets and p$+$A DY data sets,
but also firstly includes the inclusive pion production in the deutron and gold collisions at PHENIX
to constrain the nuclear gluon PDF.
The EPPS16~\cite{Eskola:2016oht} is the extension of the previous EPS09~\cite{Eskola:2009uj} with
the additional experimental data from proton-lead collisions at LHC~\cite{Chatrchyan:2014hqa,Khachatryan:2015hha,Aad:2015gta} for the first time.
It offers a less biased and flavor-dependent fitting analysis for nuclear PDFs.
Following the CTEQ global PDF fitting framework~\cite{Gao:2013xoa,Kovarik:2013sya,Dulat:2016rzo},
nCTEQ15~\cite{Kusina:2015vfa} describes the nuclear dependence of nPDFs at NLO on different nuclei, including Pb.
The nIMParton~\cite{Wang:2016mzo} is a global analysis based on two data sets of nuclear DIS data,
which either only contains isospin-scalar nuclei or all nuclear data. The difference of the fitting
functions obtained by these two data sets is on the shadowing (small x region) effects.
In addition, the Fermi motion and off-shell effect, nucleon swelling, and parton-parton recombination
are taken into account together in the nIMParton framework. In the market, some other parameterizations
of nPDFs have also been proposed~\cite{AtashbarTehrani:2012xh,Hirai:2007sx,Kulagin:2004ie,Kulagin:2014vsa,AtashbarTehrani:2017rwo}, and so far
due to the lack of enough experimental data, limited constraints and large uncertainties
appear in nuclear gluon distribution, especially at small x and large $Q^2$ region, and nuclear quark
distribution at large x for all sets of nPDFs.
%%%%%%%%%%%%%%%%%%%%%%%%%%%%%%%%%%%%%%%%%%
%%***********    Figure for isolated photon production cross section in p+p ATLAS 8TeV  and compared with Data ****
\hspace{0.7in}
\begin{figure}[!t]
\begin{center}
\hspace*{-0.1in}
\includegraphics[width=3.35in,height=2.4in,angle=0]{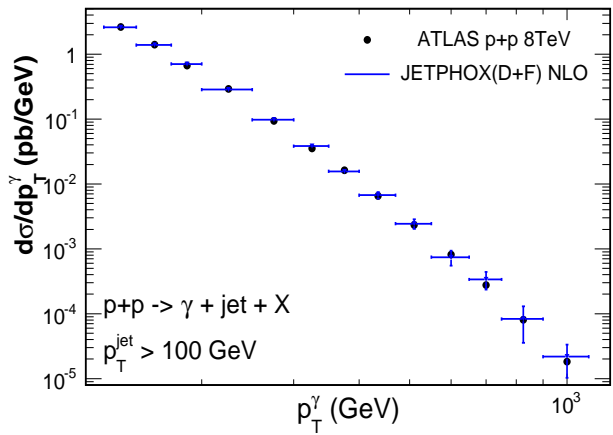}
\hspace*{-0.1in}
\caption{The cross section of isolated photon plus jet as function of $\rm p_T^{\gamma}$ in $\rm{p+p}$ collisions at
$8$~TeV and the NLO pQCD theoretical calculations(JETHOX).
}
\label{fig:ppgammapt}
\end{center}
\end{figure}
\hspace*{-1.5in}
%%%%%%%%%%%%%%%%%%%%%%%%%%%%%%%%%%%%%%%%%%%%%%%%%%%%%%%

In this work, we study the isolated prompt photon and $\gamma+$jet productions
in proton-lead collisions at LHC energy $\sqrt{\rm s_{NN}}=8.16$TeV with a NLO
pQCD program JETPHOX~\cite{Catani:2002ny,Aurenche:2006vj,Belghobsi:2009hx} with updated
proton's PDFs $-$ CT14 parametrization~\cite{Dulat:2016rzo}.
The nuclear parton distribution functions (nPDFs) parametrizations (DSSZ, EPPS16, nCTEQ15)
are performed at next-to-leading order accuracy and the nIMParton is based on Leading Order (LO)
calculation with parton-parton recombination. These four sets of nPDFs parametrizations have been
utilized in obtaining the cross sections of photon and photon associated jet in p+A collisions.
We calculate the nuclear modification factors $\rm R_{pPb}$ for isolated photon production as function of
transverse momentum $\rm p_T^{\gamma}$ and pseudo-rapidity $\eta^\gamma$ at both forward
and backward rapidity region, the forward-backward asymmetry $\rm Y_{p\rm Pb}^{asym}$
for isolated photon production as function of $\rm p_T^{\gamma}$, and the nuclear modification
factors $\rm R_{pPb}$ for $\gamma+$jet production as function of $\eta_{\gamma \rm Jet}$
at limited $\rm p_T^{\rm avg}$ intervals.
We address the quantification of dominant Bjorken x regions detected under different
specific rapidity and transverse momentum ranges.

This paper is organized as follows: in Section 2, we describe our
perturbative QCD predictions of prompt photon associated jet inclusive
cross section in proton-proton collisions at $8$TeV.
In Section 3, we discuss the nuclear modification of isolated
prompt photon productions at both forward and backward region at $8.16$TeV.
In Section 4, the cold nuclear matter effects on photon$+$jet productions are studied
at different transverse momentum intervals at $8.16$TeV.
And we give the summary in the last Section.

\section{2 Photon and Photon+Jet Productions in p+p}
\label{sec:pp}
High-$\rm p_T$ prompt photons mainly arise from two possible
mechanisms in hadronic collisions, produced directly in the hard sub-processes
referred to as "direct" photons or fragmented from energetic parton.
We consider that the next-to-leading order(NLO) inclusive cross section for the production of
prompt photon
with transverse momentum $\rm p_T^{\gamma}$ is given by the sum of the fragmentation and direct contributions,
written as~\cite{Catani:2002ny,Belghobsi:2009hx},
\begin{eqnarray}
\begin{aligned}
\sigma(p_T^{\gamma})=&\hat{\sigma}^D(p_T^{\gamma};\mu;M;M_F) \\
&+\sum_k\int_0^1\frac{dz}{z}\hat{\sigma}^F(p_T^{\gamma}/z;\mu;M;M_F)D_k^{\gamma}(z;M_F)
\end{aligned}
\label{eq:crosssection}
\end{eqnarray}
Where $\mu$ is the renormalization scale, $M$ is the initial state factorisation scale
and $M_F$ is an arbitrary final state fragmentation scale. The contribution $\hat{\sigma}^F$ denotes
the partonic cross section for producing a parton convoluted with the PDF of the incoming proton, and $D_k^{\gamma}$
is the fragmentation function of a parton $k$ (quarks,anti-quarks~and~gluon) into a photon.
$\hat{\sigma}^D$ includes the partonic cross section for producing a direct photon and the corresponding PDFs.
Experimentally, there are also secondary photons originated from hadron decay during the collisions,
therefore an isolation cut would be applied for the substantial production of photons.
A photon is isolated if the amount of deposited hadronic transverse energy
$\rm E_T$ is not more than an specific upper limit $\rm E_T^{iso}$
in a fixed radius $\rm R_{iso}=\sqrt{(\eta-\eta_\gamma)^2+(\phi-\phi_\gamma)^2}$
in pseudo-rapidity and azimuthal angle around the photon direction.
This restriction on the yields of isolated photons could not only reject the secondary decay photons,
but also reduce the contribution from fragmentation processes.
In the following we focus on the production of isolated photon and
isolated photon tagged jets in hadronic collisions.
%we apply an isolation cut to
%exterminate the background of secondary photons, as well as reduce the fragmentation contribution.
%%%%%%%%%%%%%%%%%%%%%%%%%%%%%%%%%%%%%%%%%%
%%***********    Figure for isolated photon's cold nuclear modification factor in p+Pb ATLAS 8.16TeV at forward rapidity 3-4****
\hspace{0.7in}
\begin{figure}[!t]
\begin{center}
\hspace*{-0.1in}
\includegraphics[width=3.2in,height=3.2in,angle=0]{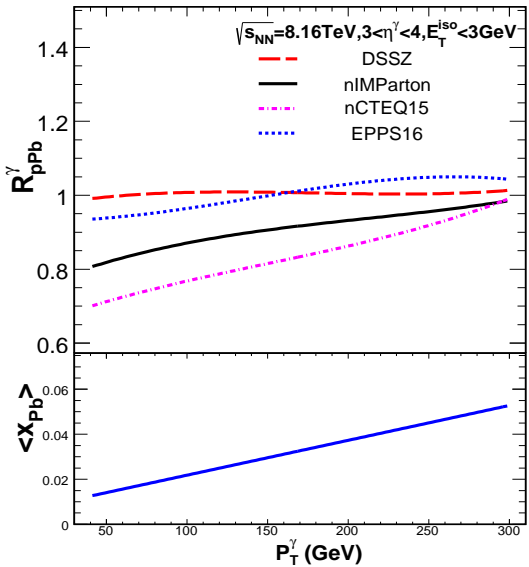}
\hspace*{-0.1in}
\caption{(Upper) A comparison between the nuclear modification ratios $\rm R_{pPb}$
for p-Pb collisions at$\sqrt{s}$ = 8.16 TeV and $\rm 3 < \eta^{\gamma} < 4$
using the nCTEQ15 , EPPS16, DSSZ and nIMParton nuclear modifications and the
CT14 free-proton PDFs. (Bottom) The corresponding average Bjorken $\rm\langle x_{Pb}\rangle$
as function of $\rm p_T^{\gamma}$.
}
\label{fig:RaF34}
\end{center}
\end{figure}
\hspace*{-1.5in}
%%%%%%%%%%%%%%%%%%%%%%%%%%%%%%%%%%%%%%%%%%%%%%%%%%%%%%%

We calculate isolated photon and jet productions in
proton-proton collisions at $\rm 8TeV$ with a next-to-leading order
pQCD program JETPHOX~\cite{Catani:2002ny,Aurenche:2006vj,Belghobsi:2009hx}
with CT14 nucleon parton distribution functions~\cite{Dulat:2016rzo}
in accordance with the ATLAS experiment~\cite{Aaboud:2016sdm}.
The isolated energy cut for a photon has been set as $\rm E_T^{iso}<6GeV$, and
the isolated cone of radius in the pseudo-rapidity and azimuthal angle plane is $\rm R_{cone}=0.4$.
Moreover photons are selected if its transverse momentum $\rm p_T^{\gamma}>130 GeV$ and $|\eta^{\gamma}|<2.37$,
except $\rm 1.37<|\eta^{\gamma}|<1.56$.
%The isolation cut not only rejects the background of secondary photons,
%it also reduces the fragmentation component.
Jets are reconstructed by anti-$\rm k_t$ algorithm with cone size $\rm R=0.6$ with $\rm p_T^{jet}>100 GeV$
and $\rm|\eta^{jet}|<4.4$.
% JetPhox theoretical study 0204023
In Fig.\ref{fig:ppgammapt}, we calculate the differential cross section
$\rm d\sigma/dp_T^{\gamma}$ up to $\rm p_T^{\gamma}=1$ TeV
in proton-proton collisions at $\rm 8$ TeV, and our theoretical
predication shows good agreement with the ATLAS experimental results.

\section{3 Isolated Photon in p+Pb collisions at very forward and backward rapidity}
\label{Nuclear modification ratio for Isolated Photon in proton$+$lead collisions at very forward and backward rapidity region}
The inclusive cross section for the isolated photon production in proton-nucleus collisions
could be evaluated by using nuclear PDFs (nPDFs) as substitutes for the free-nucleon PDFs in the collinear
factorization framework as stated above, which could effectively include different CNM effects.

In our calculations, we obtain the nPDFs $\rm f_i^A(x,Q^2)$ by multiplying the CT14 parton distribution functions (PDFs)\cite{Dulat:2016rzo}
with a flavor and scale dependent factor $\rm R_i^A(x,Q^2)$ taken from four different parametrizations
DSSZ~\cite{deFlorian:2011fp}, EPPS16~\cite{Eskola:2016oht}, nCTEQ15~\cite{Kusina:2015vfa}, nIMParton~\cite{Wang:2016mzo}.
These four parametrizations for nPDFs are similar that they categorize CNM effects with Bjorken x region into shadowing,
anti-shadowing, EMC effect and so on, but differ in the specific formalisms and parameters for describing CNM effects and the input
experimental data used in global fits. DSSZ,~nCTEQ15,~EPPS16 could be convoluted in the expression for calculating the photon
production at next-to-leading order(NLO) since they are also quantitated in the NLO pQCD framework. Whereas the leading order
results for photon production are applied with nIMParton parametrization to maintain the consistency of the analysis.
%%***********    Figure for isolated photon's cold nuclear modification factor in p+Pb ATLAS 8.16TeV at back rapidity -4~-3****
\hspace{0.7in}
\begin{figure}[!t]
\begin{center}
\hspace*{-0.1in}
\includegraphics[width=3.2in,height=3.2in,angle=0]{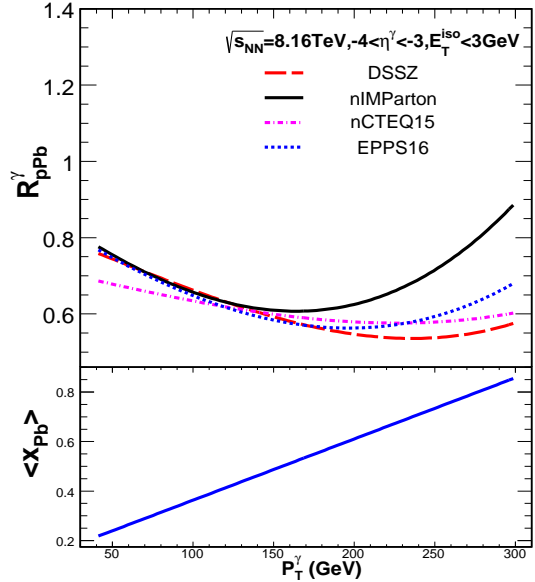}
\hspace*{-0.1in}
\caption{The same as Fig.\ref{fig:RaF34}, except at backward rapidity $\rm -4 < \eta^{\gamma} < -3$.
}
\label{fig:RaB34}
\end{center}
\end{figure}
\hspace*{-1.5in}
%%%%%%%%%%%%%%%%%%%%%%%%%%%%%%%%%%%%%%%%%%%%%%%%%%%%%%%
%%***********    Figure for isolated photon's cold nuclear modification factor in p+Pb ATLAS 8.16TeV at back rapidity -4~-3****
\hspace{0.7in}
\begin{figure*}[!t]
\begin{center}
\hspace*{-0.1in}
\includegraphics[width=3.2in,height=2.2in,angle=0]{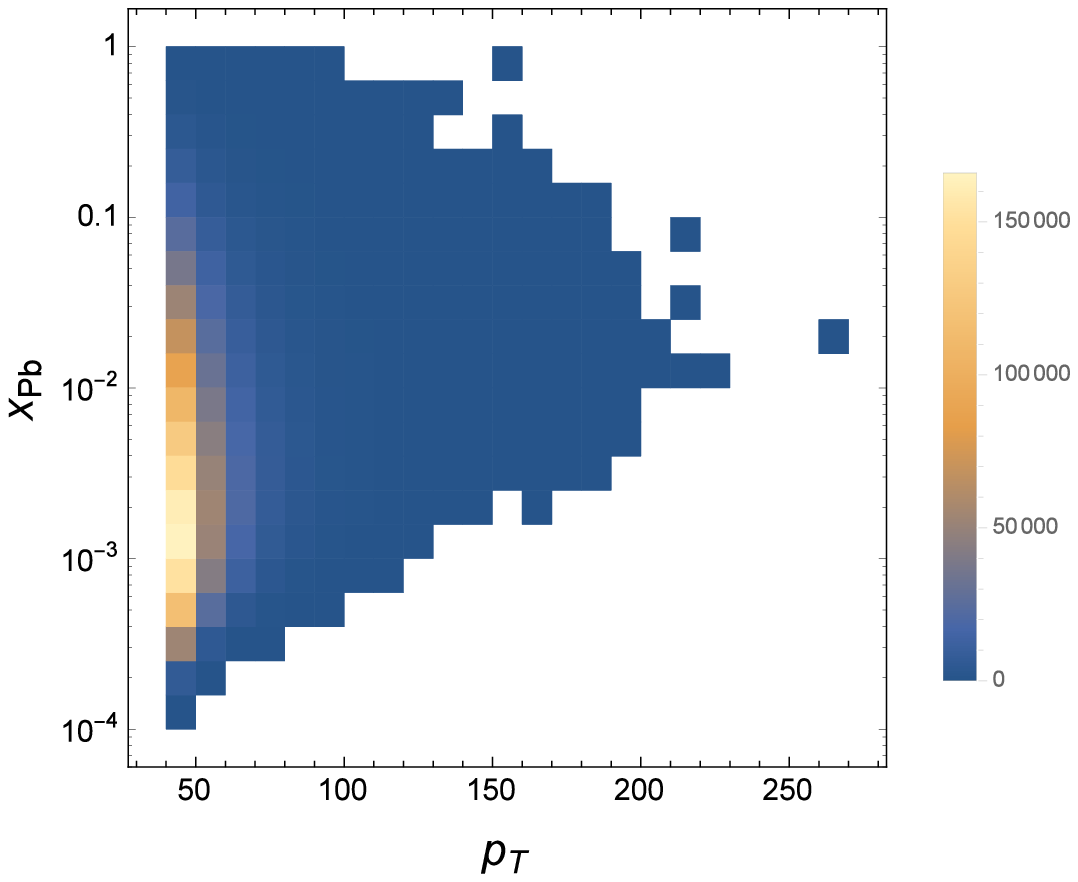}
\includegraphics[width=3.2in,height=2.2in,angle=0]{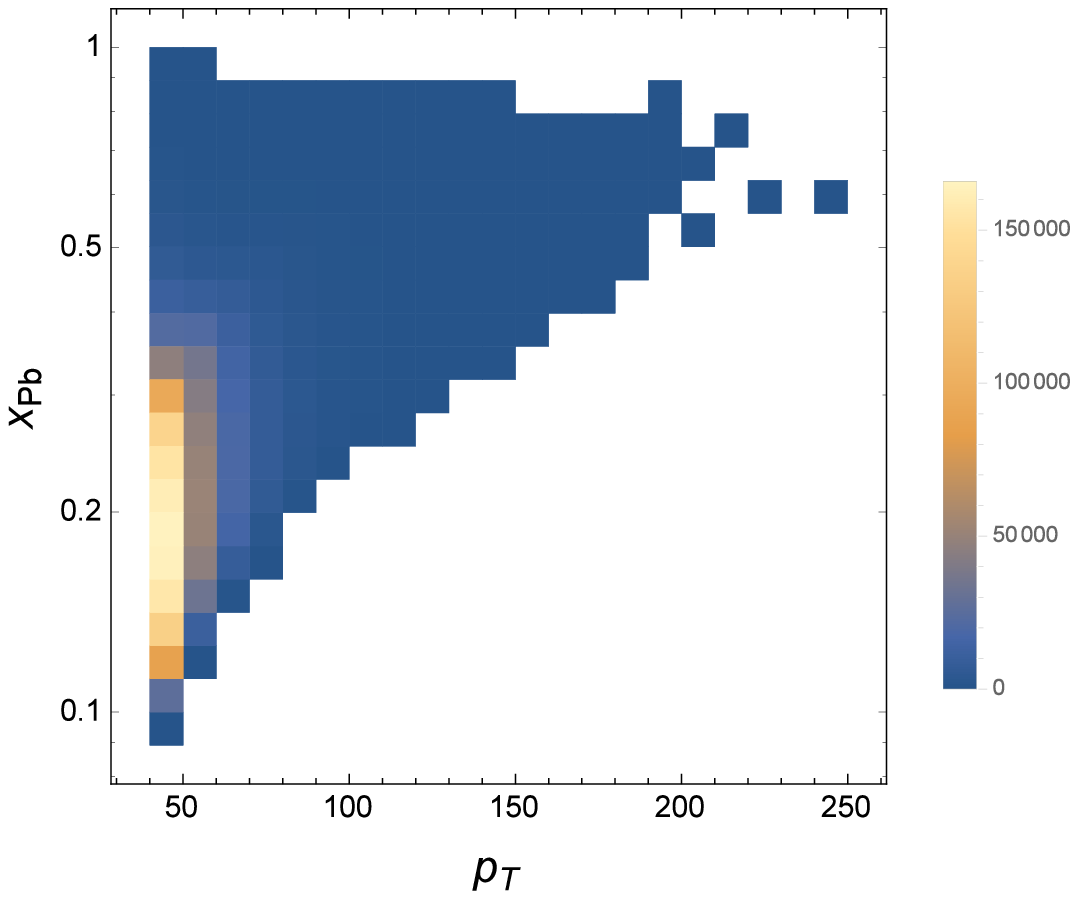}
\hspace*{-0.1in}
\caption{(Left)NLO fluctuations at Forward rapidity $\rm 3<\eta^{\gamma}<4$;
(Right)NLO fluctuations at Backward rapidity $-4<\eta^{\gamma}<-3$.
}
\label{fig:fbcorr}
\end{center}
\end{figure*}
\hspace*{-1.5in}
%%%%%%%%%%%%%%%%%%%%%%%%%%%%%%%%%%%%%%%%%%%%%%%%%%%%%%%

The nuclear modification factors in proton+lead collisions are defined as:
\begin{eqnarray}
\rm R_{pPb}=\frac{d\sigma^{pPb}/dp_T}{\langle N_{coll}\rangle d\sigma^{pp}/dp_T}
\label{eq:Rab}
\end{eqnarray}
with $\rm\langle N_{coll}\rangle$ representing the number of binary nucleon-nucleon collisions by the glauber model~\cite{dEnterria:2003xac}.

Now we can make our theoretical predictions for the isolated prompt photon
production in p$+$p and p$+$Pb collisions at very forward rapidity region
$\rm 3<\eta^{\gamma}<4$ at $\rm \sqrt{s_{NN}}=8.16$TeV with ATLAS isolated cuts for photons~\cite{Aaboud:2016sdm},
along with photon's transverse momentum constrained in $\rm 40 GeV<p_T^{\gamma}<300 GeV$.
We display the nuclear modification ratio $\rm R_{pPb}^{\gamma}$ as function of $\rm p_T^{\gamma}$ in upper panel of Fig.\ref{fig:RaF34}.
And the momentum fraction carried by initial parton from the incoming particle can be roughly estimated at LO as
$\rm x_{1,2}=\frac{p_T}{\sqrt{s_{NN}}}(e^{\pm y_1}+e^{\pm y_2})$, which $\rm x_1(x_p)$ is the initial parton
coming from proton on the $\rm +z$ direction, $\rm x_2(x_{pb})$ is the initial parton coming from lead on the
$\rm -z$ direction in $\rm p+Pb$ collisions and $y_{1,2}$ is the rapidity of $\gamma$ and the associated jet respectively.
The estimated average Bjorken $\rm\langle x_{Pb}\rangle$ has been defined as the events average value of Bjorken $\rm x_{Pb}$
in JetPhox simulation.
In the bottom panel of Fig.\ref{fig:RaF34}, we show the estimation of the parton's average momentum fraction off nucleus based on NLO results in JETPHOX. We have checked that $\rm\langle x_{Pb}\rangle$ for nPDFs parametrizations vary slightly from each other,
which $\rm\langle x_{Pb}\rangle$ for nIMParton can be calculated directly at LO.
We can see that the average Bjorken $\rm\langle x_{Pb}\rangle$ is lower than 0.055 at very forward rapidity region,
which represents the shadowing effect dominating the CNM effects. Moreover the average Bjorken $\rm\langle x_{Pb}\rangle$ has a linear
positive dependence with $\rm p_T^{\gamma}$ expected in its LO estimation.

In the Fig.\ref{fig:RaF34}, we see that the DSSZ's shadowing effect is unremarkable on the suppression
of isolation photon production in p$+$Pb collisions when the average Bjorken $\rm\langle x_{Pb}\rangle<0.055$.
We also notice that DSSZ's $\rm R_{pPb}^{\gamma}(p_T^{\gamma})$ shows a very weak $\rm p_T^{\gamma}$ dependence,
which means its shadowing effect is nearly independent on photon's transverse momentum in DSSZ at forward rapidity region.
Meanwhile the other three parametrizations' shadowing decrease with $\rm p_T^{\gamma}$ increasing upon $\rm 3<\eta^{\gamma}<4$.
Also we can find that the brand new nPDF parametrization nIMParton's shadowing is only weaker than nCTEQ15's in our discussion.
%%%%%%%%%%%%%%%%%%%%%%%%%%%%%%%%%%%%%%%%%%
%%***********    Figure for isolated photon's forward-to-backward asymmetry in p+Pb ATLAS 8.16TeV at forward rapidity 3-4****
\hspace{0.7in}
\begin{figure}[!t]
\begin{center}
\hspace*{-0.1in}
\includegraphics[width=3.2in,height=2.4in,angle=0]{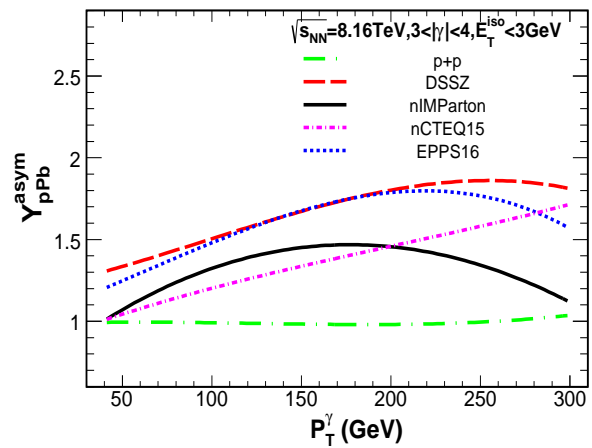}
\hspace*{-0.1in}
\caption{A comparison between the forward-to-backward asymmetry $\rm Y_{asym}$
for p-Pb collisions at$\sqrt{s_{NN}}$ = 8.16 TeV and $\rm 3 < \eta^{\gamma} < 4$
using the nCTEQ15 , EPPS16, DSSZ and nIMParton nuclear modifications and the
CT14 free-proton PDFs.
}
\label{fig:asym34}
\end{center}
\end{figure}
\hspace*{-1.5in}
%%%%%%%%%%%%%%%%%%%%%%%%%%%%%%%%%%%%%%%%%%%%%%%%%%%%%%%

In Fig.\ref{fig:RaB34}, similar phenomenon that the positively linear correlation
between $\rm\langle x_{Pb}\rangle$ and $\rm p_T^{\gamma}$ has been shown at backward rapidity $-4<\eta^{\gamma}<-3$.
Whereas the estimated average Bjorken $\rm\langle x_{Pb}\rangle$ ranges from 0.25 to 0.8,
which mostly correspond to EMC effect. We could go a little further to distinguish
four different parametrizations' EMC maximum from each nuclear modification factor's extreme point,
such as the nIMParton's EMC minimum appears in $\rm p_T^{\gamma}=150GeV$ and the corresponding
average Bjorken locates around $\rm\langle x_{\rm Pb}\rangle=0.5$, which is the lowest in our results.
We further investigate the correlations between $\rm x_{Pb}$ and $\rm p_T^{\gamma}$ at both forward and backward
rapidities at NLO, shown in Fig.~\ref{fig:fbcorr}.
We observe the broadening of Bjorken $\rm x_{Pb}$ at specific $\rm p_T^{\gamma}$ interval due to higher corrections,
and the spreading of  $\rm x_{Pb}$ at small $\rm p_T^{\gamma}$ is rather wider.
Also we can see a very dense statistics cluster around the low $\rm p_T^{\gamma}$ in our Monte-Carlo simulation,
because the possibility distribution of the hard sub-processes for the photon production following
double-logarithmic declining with photon's transverse momentum.

Note the nuclear modification factor is sensitive to the nucleon
PDF(nPDF) and p+p baseline~\cite{CMS:2013cka}, we may calculate the ratio
of the photon production at forward and backward rapidity, which could get rid of
the large uncertainty in free nucleon PDFs, which could be used to probe the CNM effects with less
arbitrariness\cite{Helenius:2014qla,Goharipour:2017uic,Goharipour:2018sip}.
We define the forward-backward yield asymmetry as:
\begin{eqnarray}
\rm Y_{p\rm Pb}^{asym}=\frac{d\sigma/dp_T(p+\rm Pb \rightarrow \gamma+X)|_{\eta\in[\eta_1,\eta_2]}}{d\sigma/dp_T(p+\rm Pb \rightarrow \gamma+X)|_{\eta\in[-\eta_2,-\eta_1]}}
\label{eq:asym}
\end{eqnarray}

Our predictions of the forward-to-backward yield asymmetries $\rm Y_{p\rm Pb}^{asym}$
for the isolated prompt photon production in p$+$Pb collisions at $\rm \sqrt{s}=8.16TeV$
and $\rm 3 < |\eta^{\gamma}| < 4$ are shown in Fig.\ref{fig:asym34}.
As a result of the symmetry of the colliding system, there is nearly no forward-to-backward
yield asymmetry observed in proton-proton collisions.
$\rm Y_{p\rm Pb}^{asym}$ is larger than one in p$+$Pb collisions,
which means the photon production suffering more suppression in the backward rapidity
region.  And the EMC effect reduces the photon production more effectively
than the shadowing effect does on the whole.
Besides, we notice that the value of $\rm Y_{p\rm Pb}^{asym}$ starts going down to one
with all parametrizations due to the decreasing of EMC effect at relatively high $\rm p_T^{\gamma}$.
This manifestation seems less obvious in nCTEQ15 as its EMC maximum is close to $\rm p_T^{\gamma}$'s highest boundary
and our approximate curve fitting.
%%%%%%%%%%%%%%%%%%%%%%%%%%%%%%%%%%%%%%%%%%
%%***********    Figure for isolated photon's cold nuclear modification factor in p+Pb ATLAS 8.16TeV at forward rapidity 3-4****
\hspace{0.7in}
\begin{figure}[!t]
\begin{center}
\hspace*{-0.1in}
\includegraphics[width=3.2in,height=3.2in,angle=0]{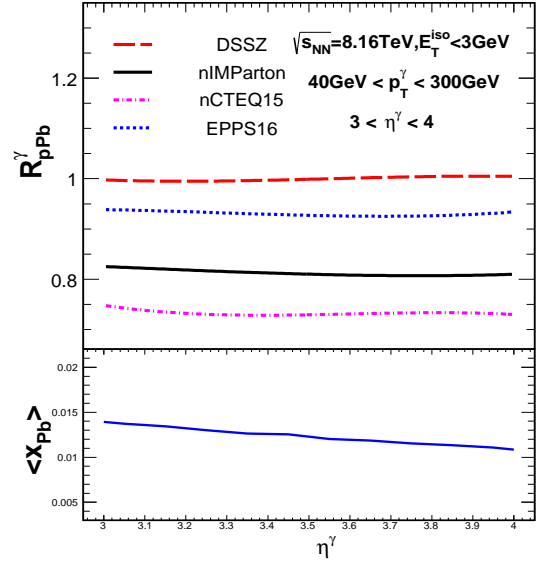}
\hspace*{-0.1in}
\caption{The same as Fig.\ref{fig:RaF34}, but as a function of photon's rapidity $\eta^{\gamma}$.
}
\label{fig:ReF34}
\end{center}
\end{figure}
\hspace*{-1.5in}
%%%%%%%%%%%%%%%%%%%%%%%%%%%%%%%%%%%%%%%%%%%%%%%%%%%%%%%
%%%%%%%%%%%%%%%%%%%%%%%%%%%%%%%%%%%%%%%%%
%%***********    Figure for isolated photon's cold nuclear modification factor in p+Pb ATLAS 8.16TeV at forward rapidity 3-4****
\hspace{0.7in}
\begin{figure}[!t]
\begin{center}
\hspace*{-0.1in}
\includegraphics[width=3.2in,height=3.2in,angle=0]{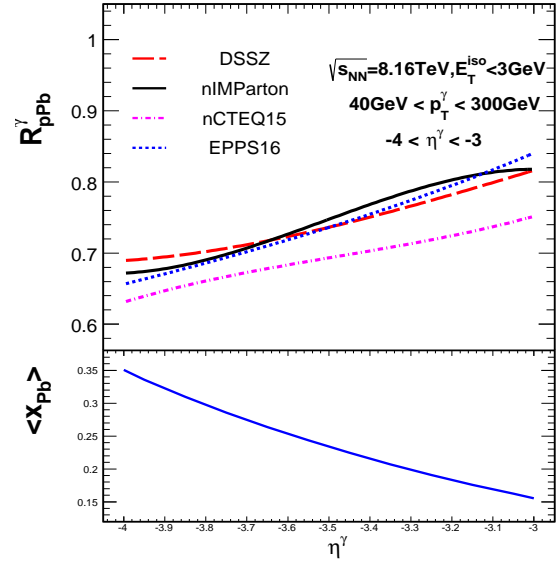}
\hspace*{-0.1in}
\caption{The same as Fig.\ref{fig:RaB34}, but as a function of photon's rapidity $\eta^{\gamma}$.
}
\label{fig:ReB34}
\end{center}
\end{figure}
\hspace*{-1.5in}
%%%%%%%%%%%%%%%%%%%%%%%%%%%%%%%%%%%%%%%%%%%%%%%%%%%%%%%

In order to further explore the impact of input nuclear modifications
on the cross section of isolated prompt photon productions in proton-nucleus collision.
We further discuss the isolated photon's nuclear modification factor $\rm R_{pPb}^{\gamma}(\eta^{\gamma})$
as a function of photon's rapidity $\eta^{\gamma}$ at both forward and
backward rapidities. The Fig.\ref{fig:ReF34} tells us that a growing suppression
on the photon productions from DSSZ, EPPS16, nIMParton and nCTEQ15 at forward pseudo-rapidity,
which quantitatively appears in accordance with the $\rm R_{pPb}^{\gamma}(p_T^{\gamma})$ at
$p_T^{\gamma}=50 GeV$ due to the highest statistics at lowest $p_T^{\gamma}$ region in Monte-Carlo
simulations exhibited above in Fig.\ref{fig:RaF34}. On the other hand, $\rm R_{pPb}^{\gamma}(\eta^{\gamma})$
shows very weak $\eta^{\gamma}$ dependence, because the variation on Bjorken $\rm x_{Pb}$ is at the magnitude
of $10^{-3}$ at region $3<\eta^{\gamma}<4$, shown in the bottom of Fig.\ref{fig:ReF34}.
Combining the results of $\rm R_{pPb}$ evolved with $p_T^{\gamma}$ and $\eta^{\gamma}$, the suppression pattern of isolated
photon could be quantitatively analyzed through $\langle x_{Pb}\rangle$ at both forward and backward rapidities.
In Fig.\ref{fig:ReB34}, the nuclear modification factors using four different nPDFs all show the nearly positive linear relation
with $\eta^{\gamma}$, which the values could match with $\rm R_{pPb}^{\gamma}(p_T^{\gamma})$'s through the average Bjorken
$\langle x_{Pb}\rangle$, shown in Fig.\ref{fig:ReB34} and Fig.\ref{fig:RaB34} respectively.
And the nCTEQ15 parametrization gives a stronger suppression than the others', which could be
confirmed in the prediction of $\rm R_{pPb}^{\gamma}(p_T^{\gamma})$ at low $\rm p_T^{\gamma}$ region in Fig.\ref{fig:RaB34}.

%%%%%%%%%%%%%%%%%%%%%%%%%%%%%%%%%%%%%%%%%%
%%***********    Figure for isolated photon plus jet's correlations at NLO****
\hspace{0.7in}
\begin{figure*}[!t]
\begin{center}
\hspace*{-0.1in}
\includegraphics[width=2.25in,height=1.85in,angle=0]{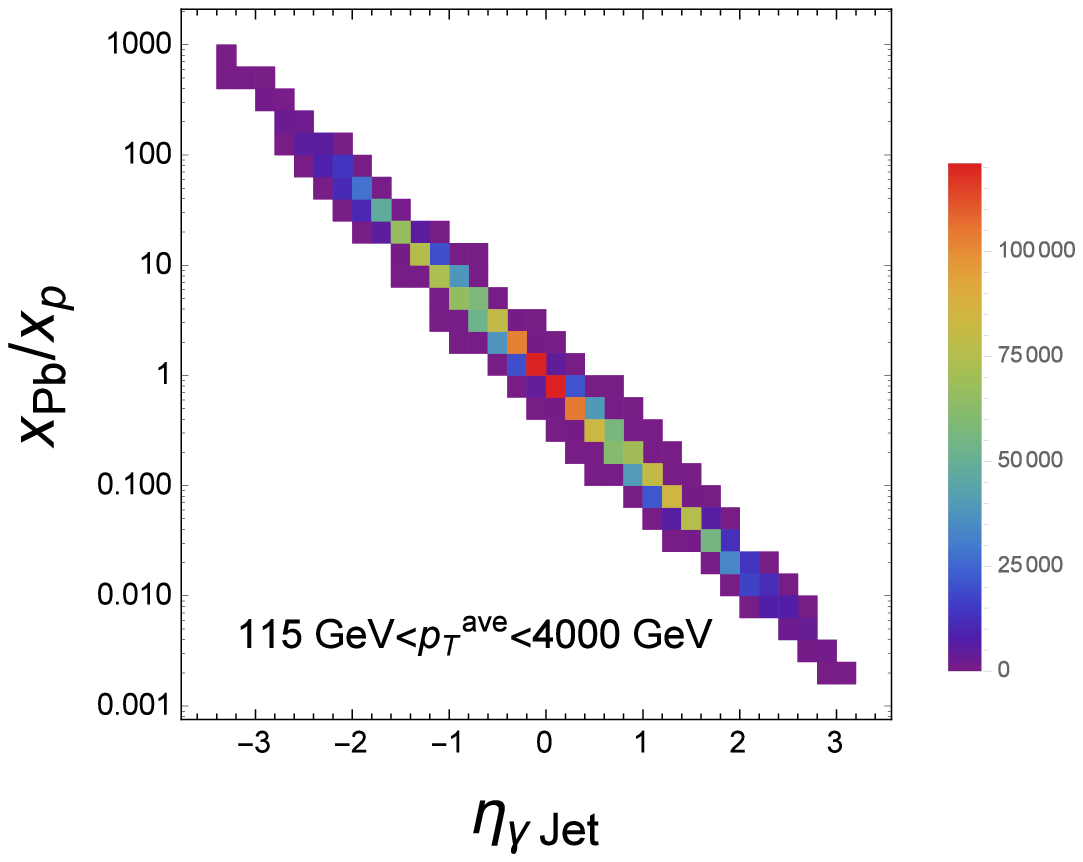}
\includegraphics[width=2.25in,height=1.85in,angle=0]{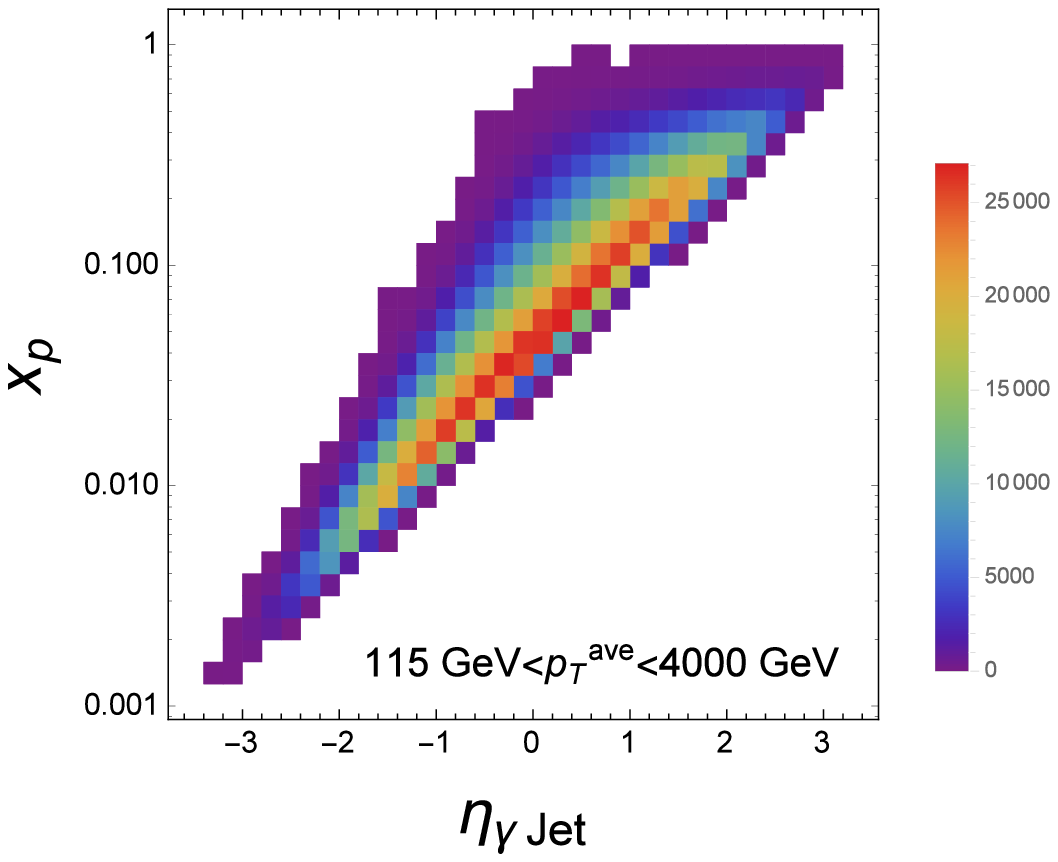}
\includegraphics[width=2.25in,height=1.85in,angle=0]{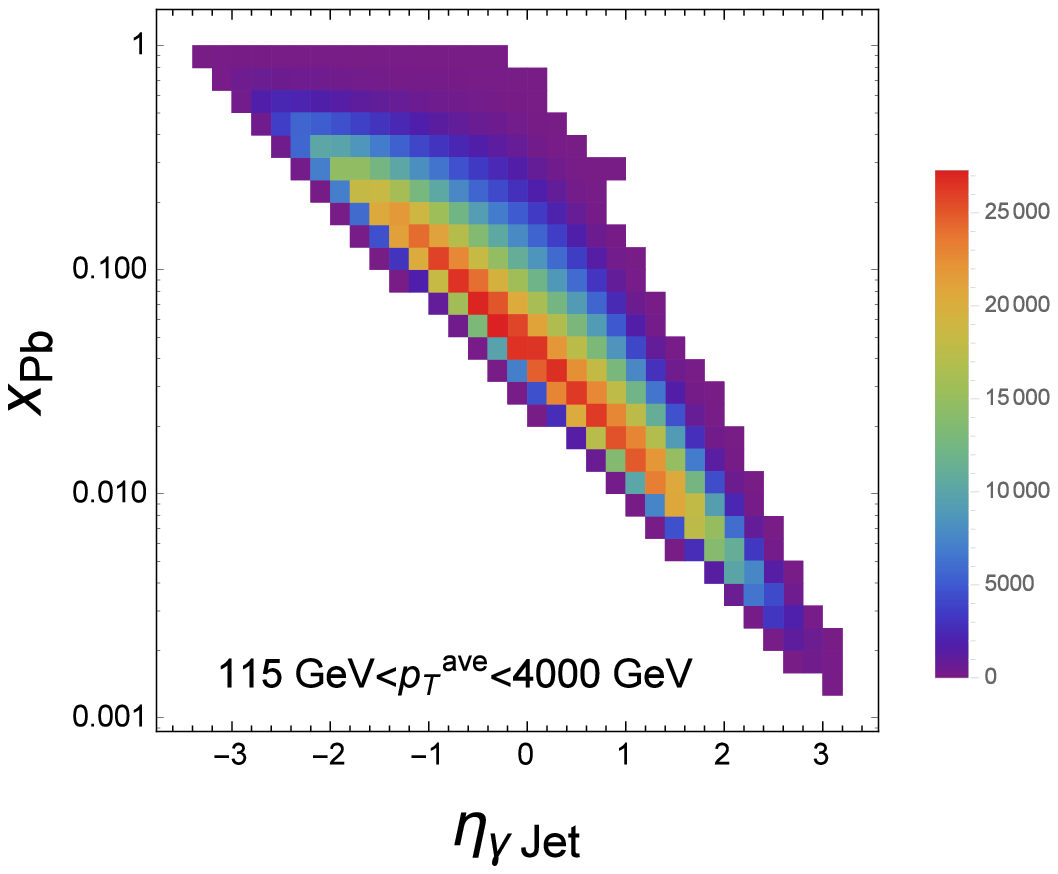}
\hspace*{-0.1in}
\caption{Correlation between $\rm x_{Pb}/x_p$(Left), $x_p$(Middle), $x_{Pb}$(Right) and $\gamma+$jet pseudo-rapidity $\rm \eta_{\gamma jet}$.
}
\label{fig:correlation1}
\end{center}
\end{figure*}
\hspace*{-1.0in}
%%%%%%%%%%%%%%%%%%%%%%%%%%%%%%%%%%%%%%%%%%%%%%%%%%%%%%%

\section{4 Isolated Photon+Jet in p+Pb}
\label{Isolated Photon associated with Jet Productions in proton$+$lead collisions}

As compared to isolated photon productions, the isolated photon associated jet production in p+A reactions has more leverage power to access CNM effects in a wider kinematic regions due to its exclusive property.
To understand the nuclear modifications for isolated prompt photon associated jet productions,
one usually estimates the momentum fractions of the initial-state partons at leading order(LO)
to evaluate the CNM effects' contribution by the final-state kinematics,
namely $\rm x_{1,2}$ defined above. In the following,
we are enlightened by the work on dijets productions in the CMS collaboration~\cite{Sirunyan:2018qel},
and provide the $\gamma$+jet pseudo-rapidity $\rm \eta_{\gamma jet}=\frac{1}{2}(\eta_\gamma+\eta_{Jet})$ distributions
of a photon tagged jet at specific range of their average transverse momentum $\rm p_T^{avg}=\frac{1}{2}(p_T^{\gamma}+p_T^{Jet})$.
As $\rm \eta_{\gamma jet}$ would be equal to $\rm \frac{1}{2}\ln(x_p/x_{Pb})$ in
the center-of-mass frame when two partons collide with each other at LO.
The NLO simulation would give rise to complicated correlations between $\rm x_{Pb}$ ($\rm x_{p}$) and $\rm \eta_{\gamma jet}$  describing the nuclear matter's influence,
shown in  Fig.~\ref{fig:correlation1}.
We also compute the correlations between $\rm x_{Pb}/x_p$ and $\rm \eta_{\gamma Jet}$
at $\rm 115GeV<p_T^{avg}<4000GeV$ interval in Fig.~\ref{fig:correlation1}. It is shown though at NLO accuracy, both distributions of $\rm x_{Pb}$ and
$\rm x_{p}$ over $\rm \eta_{\gamma jet}$ are rather wider, when the ratio $\rm x_{Pb}/x_p$ at NLO is rather narrow and centered at values at LO with very high statistics.
%%%%%%%%%%%%%%%%%%%%%%%%%%%%%%%%%%%%%%%%%%
%%***********    Figure for isolated photon plus jet's correlations at NLO****
\hspace{0.7in}
\begin{figure}[!t]
\begin{center}
\hspace*{-0.1in}
\includegraphics[width=3.2in,height=2.2in,angle=0]{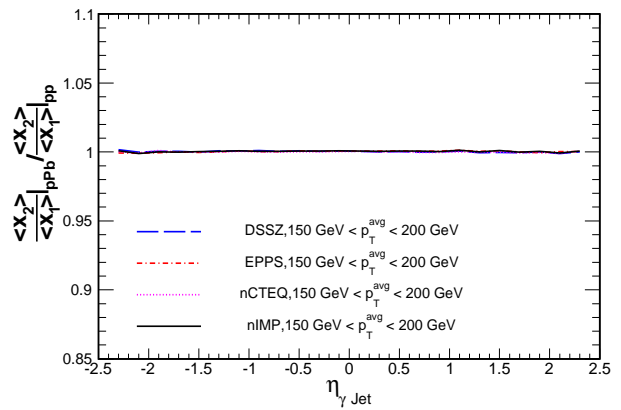}
\hspace*{-0.1in}
\caption{The ratio of $\langle x_2\rangle/\langle x_1\rangle$ in p$+$Pb and p$+$p collisions
as a function of $\rm\eta_{\gamma jet}$, considering different CNM effects descriptions.
}
\label{fig:diff}
\end{center}
\end{figure}
\hspace*{-1.0in}
%%%%%%%%%%%%%%%%%%%%%%%%%%%%%%%%%%%%%%%%%%%%%%%%%%%%%%%

\begin{figure}[!t]
\begin{center}
\hspace*{-0.1in}
\includegraphics[width=3.2in,height=2.2in,angle=0]{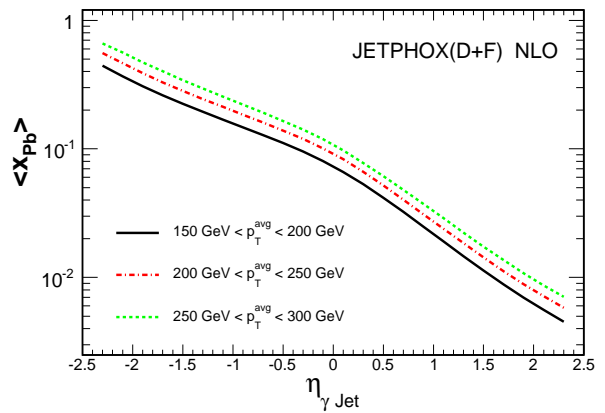}
\hspace*{-0.1in}
\caption{Mean Bjorken x of the parton from the lead ion $\rm x_{Pb}$ obtained from JetPhox
as a function of $\rm\eta_{\gamma jet}$ in different $\gamma+$jet events' $\rm p_T^{avg}$ intervals.
}
\label{fig:correlation2}
\end{center}
\end{figure}

In Fig.\ref{fig:diff}, we can notice that there are tiny shifts on $\frac{<x_2>}{<x_1>}$ as a function of $\rm\eta_{\gamma jet}$
caused by nuclear matter. Based on the relation between $\rm \eta_{\gamma Jet}$ and $\rm\langle x_{Pb}\rangle$,
we could reach an assessment about different CNM effects predominate region at $\rm \eta_{\gamma Jet}$,
which the $\gamma+$jet production is sensitive to shadowing ($\rm \eta_{\gamma Jet}>1.6$),
anti-shadowing (\rm $-0.2<\eta_{\gamma Jet}<1.6$), and EMC effects ($\rm \eta_{\gamma Jet}<-0.2$)
at $\rm 150<p_T^{avg}<200$ interval, shown as the black line in Fig.\ref{fig:correlation2}.
The average Bjorken $\rm\langle x_{Pb}\rangle$ shows a nearly negative log-linear relation with
$\rm \eta_{\gamma Jet}$, and becomes globally higher when the $\rm p_T^{avg}$ interval increasing.
%%%%%%%%%%%%%%%%%%%%%%%%%%%%%%%%%%%%%%%%%%
%%***********    Figure for isolated photon plus jet's correlations at NLO****
\hspace{0.7in}
\hspace*{-1.0in}
%%%%%%%%%%%%%%%%%%%%%%%%%%%%%%%%%%%%%%%%%%%%%%%%%%%%%%%

Furthermore, we discuss the nuclear modification factor for $\gamma+$jet production
as a function of $\rm \eta_{\gamma jet}$ at $\rm 115GeV<p_T^{avg}<4000GeV$ and
$\rm 200GeV<p_T^{avg}<250GeV$. We have roughly estimated that the nuclear modification
factor is sensitive to shadowing when $\rm \eta_{\gamma jet}>1.6$
and is sensitive to EMC effect when $\rm \eta_{\gamma jet}<-0.2$.
In the total average transverse momentum interval ($\rm 115GeV<p_T^{avg}<4000GeV$),
we could find that nIMParton offers the strongest shadowing effect, and EPPS16 has the most
predominant anti-shadowing effect and EMC effect, along with DSSZ's is in between these two.
At the meantime, nCTEQ15 provides a almost constant suppression when $\rm \eta_{\gamma jet}$ varies.
When the average transverse momentum interval limited to $\rm 200GeV<p_T^{avg}<250GeV$,
nCTEQ15's shadowing starts to restore, as well as it remains flattened at small $\rm \eta_{\gamma jet}$
and large average Bjorken $\rm \langle x_{Pb}\rangle$. nIMParton's and EPPS16's shadowing becomes more damped.
And EPPS16 has more clear anti-shadowing peak as well as nIMParton and DSSZ begin to reveal anti-shadowing peak.
DSSZ provides the strongest EMC suppression at this situation. It is emphasized by leveraging the rapidity and
transverse momentum of both photon and jet with measured $\rm p_T^{avg}$ and $\rm \eta_{\gamma jet}$,
we can get access to CNM effects in a wide regime and also allocate different kinematics precisely, where
differences between varieties of nPDFs sets may be investigated more effectively.
%%%%%%%%%%%%%%%%%%%%%%%%%%%%%%%%%%%%%%%%%%
%%***********    Figure for isolated photon plus jet's RpA****
\hspace{0.7in}
\begin{figure}[!t]
\begin{center}
\hspace*{-0.1in}
\includegraphics[width=3.3in,height=2.8in,angle=0]{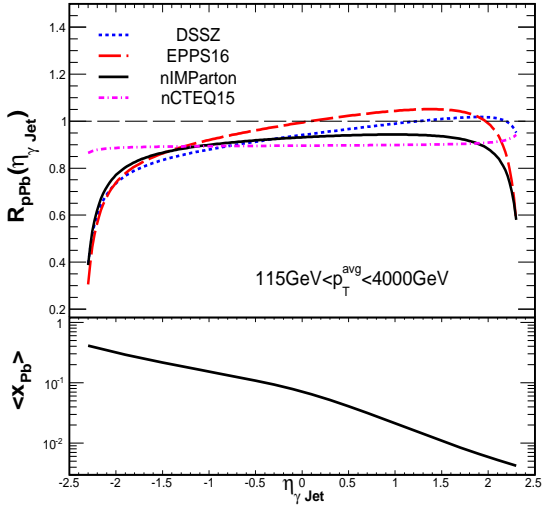}
\hspace*{-0.1in}
\caption{(Upper) The nuclear modification factor for $\gamma +$ jet production as a function of $\rm\eta_{\gamma jet}$
at $\rm 115 GeV<p_T^{avg}<4000 GeV$. The NLO pQCD calculation is based on JetPhox with nCTEQ15 , EPPS16, DSSZ and nIMParton
as the nPDFs. (Bottom) The corresponding average Bjorken $\rm\langle x_{Pb}\rangle$
as function of $\rm \eta^{\gamma Jet}$.
}
\label{fig:RpA0}
\end{center}
\end{figure}
\hspace*{-1.5in}
%%%%%%%%%%%%%%%%%%%%%%%%%%%%%%%%%%%%%%%%%%%%%%%%%%%%%%%
%%%%%%%%%%%%%%%%%%%%%%%%%%%%%%%%%%%%%%%%%%
%%***********    Figure for isolated photon plus jet's RpA****
\hspace{0.7in}
\begin{figure}[!t]
\begin{center}
\hspace*{-0.1in}
\includegraphics[width=3.3in,height=2.8in,angle=0]{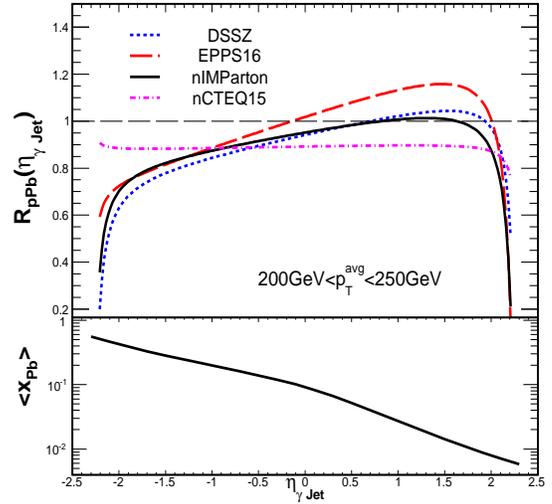}
\hspace*{-0.1in}
\caption{The same as Fig.\ref{fig:RpA0},except for at $\rm 200 GeV<p_T^{avg}<250 GeV$
}
\label{fig:RpA2}
\end{center}
\end{figure}
\hspace*{-1.5in}
%%%%%%%%%%%%%%%%%%%%%%%%%%%%%%%%%%%%%%%%%%%%%%%%%%%%%%%
\section{5 Summary}
\label{Summary and Conclusions}
%discuss the impact of CNM effects on the shift of rapidity
We calculate the productions of isolated prompt photon and $\gamma +$ jet
 in p+A with CNM effects from four sets of nuclear parton
distribution functions (nPDFs) parametrizations, {\it i.e.} DSSZ, EPPS16, nCTEQ15, nIMParton,
by utilizing the NLO pQCD approach(nIMParton at LO) at LHC $8.16$ TeV. We present the nuclear modification ratio of
isolated prompt photon $\rm R_{pA}^{\gamma}(p_T^{\gamma})$ and $\rm R_{pA}^{\gamma}(\eta^{\gamma})$,
and find shadowing effect and EMC effects dominating at very forward and backward rapidity region respectively.
And the rapidity dependence of prompt photon's nuclear modification ratio
shows weak rapidity dependence at forward region and varies linearly at backward rapidity region.
The production of the isolated photon associated with jet gives us the leveraged power to study
the tomography of cold nuclear matter.
And the CNM effects of $\gamma$+jet productions could be intuitively presented at
constraint rapidity $\rm\eta_{\gamma jet}$ and average transverse momentum $\rm p_T^{avg}$ region in our discussion.
In terms of sensitivity, comparisons of four different nPDFs parametrization's cold nuclear matter
contribution have been exhibited, and nCTEQ15 shows peculiar results than the others,
which could extend our understanding on the constraints of different nPDFs descriptions.
It is noted in our work the nPDF parametric form (nIMParton) proposed by Institute of Modern Physics in China
has been applied for the first time to investigate hard processes in ultra-relativistic heavy-ion collisions,
with a comparison again other three mainstream nPDF groups' predictions.

{\bf Acknowledgments:} This research is supported by the NSFC of China with Project
Nos. 11435004, 11322546, 11805167
and partly supported by China University of Geosciences
(Wuhan) (No. 162301182691). .

\end{document}